\documentclass[twocolumn,english,aps]{revtex4}
\usepackage[T1]{fontenc}
\usepackage[latin9]{inputenc}
\usepackage{amssymb}
\usepackage{graphicx}
\usepackage{esint}

\makeatletter
\@ifundefined{textcolor}{}
{%
 \definecolor{BLACK}{gray}{0}
 \definecolor{WHITE}{gray}{1}
 \definecolor{RED}{rgb}{1,0,0}
 \definecolor{GREEN}{rgb}{0,1,0}
 \definecolor{BLUE}{rgb}{0,0,1}
 \definecolor{CYAN}{cmyk}{1,0,0,0}
 \definecolor{MAGENTA}{cmyk}{0,1,0,0}
 \definecolor{YELLOW}{cmyk}{0,0,1,0}
 }

\makeatother

\usepackage{babel}
\begin{document}

\title{Charged Impurity Scattering in Graphene Nanostructures}

\author{Zhun-Yong Ong}

\email{zhunyong.ong@utdallas.edu}

\selectlanguage{english}%

\affiliation{Department of Materials Science and Engineering, University of Texas at Dallas RL10, 800 W Campbell Rd RL10, Richardson, TX 75080}

\author{Massimo V. Fischetti}

\email{max.fischetti@utdallas.edu}

\selectlanguage{english}%

\affiliation{Department of Materials Science and Engineering, University of Texas at Dallas RL10, 800 W Campbell Rd RL10, Richardson, TX 75080}

\begin{abstract}
We study charged impurity scattering and static screening in a top-gated substrate-supported graphene nanostructure. Our model describes how boundary conditions can be incorporated into scattering, sheds light on the dielectric response of these nanostructures, provides insights into the effect of the top gate on impurity scattering, and predicts that the carrier mobility in such graphene heterostructures decreases with increasing top dielectric thickness and higher carrier density. An increase of up to almost 60 percent in carrier mobility in ultrathin top-gated graphene is predicted.
\end{abstract}
\maketitle

{\it Introduction--}
The electron mobility in single-layer graphene (SLG) has been demonstrated to be as high as 200,000 $\mathrm{cm}^{2}\mathrm{V}^{-1}\mathrm{s}^{-1}$~\cite{KIBolotin:SSC08}, and is central to many of its potential nanoelectronic applications. However, in most graphene-based heterostructures, SLG must be
physically supported by an insulating dielectric substrate such as $\mathrm{SiO}_{2}$, and the carrier mobility in such structures is about one order of magnitude lower \cite{KIBolotin:SSC08} as a result of scattering with impurities, defects, and surface roughness. This reduction in carrier mobility can be further exacerbated when a top gate, consisting of a layer of high-$\kappa$ dielectric material such as $\mathrm{HfO}_{2}$ or $\mathrm{Al_{2}}\mathrm{O}_{3}$ overlayed with metal, is deposited on SLG~\cite{MCLemme:SSE08,JSMoon:EDL10,JPezoldt:PSS10}.

Experiments suggest that the dominant factor limiting electrical transport is scattering by charged impurities, which are believed to be located at or near the SLG-substrate interface~\cite{BFallahazad:APL10}. Adam and co-workers have successfully explained the linearity of the conductivity with respect to carrier density as a consequence of charged impurity scattering~\cite{SAdam:SSC09,SAdam:PNAS07}. However, their theory is limited to the simple geometry of SLG supported by a substrate. In more realistic graphene heterostructures, the SLG is encapsulated between the substrate and the top gate~\cite{BFallahazad:APL10,BFallahazad:APL12,JARobinson:ACSNano10}, an arrangement which offers better local electrostatic control than the bottom gate and is probably required for large scale integration. On the other hand, top gates modify the dielectric environment of the SLG by introducing image charges along the various interfaces. Thus, given the significance of impurity scattering, it is important to treat this process theoretically by accounting fully for these geometrical effects in order to optimize the design of SLG-based nanoelectronics.

Jena and Konar~\cite{DJena:PRL07} suggest that charged impurity scattering in low-dimensional semiconductor nanostructures can be damped by coating them with high-$\kappa$ dielectrics.
It has been shown~\cite{CJang:PRL08,AKMNewaz:Nature12} that placing a higher-$\kappa$ overlayer ({\em e.g.}, ice in Ref.~\cite{CJang:PRL08}, non-polar liquid in Ref.~\cite{AKMNewaz:Nature12}) on SLG can lead to a weakening of the Coulombic scattering forces and can lead to an improvement in the conductivity (although Ponomarenko and co-workers using a high-$\kappa$ liquid overlayer were not able find any significant improvement~\cite{LAPonomarenko:PRL09}). The deposition of dielectric films, such as $\mathrm{HfO}_{2}$ or $\mathrm{Al_{2}}\mathrm{O}_{3}$, on graphene has been found to lead to the degradation of electron mobility~\cite{NYGarces:JAP11,MCLemme:SSE08}. This has been attributed to the roughening of the graphene surface. However, using atomic layer deposition (ALD), Fallahazad and co-workers~\cite{BFallahazad:APL10,BFallahazad:APL12} have been able to deposit ultrathin high-$\kappa$ dielectric materials on SLG with much less roughness and observe a significant improvement in the carrier mobility. Hollander and co-workers were also able to see an increase of the Hall mobility in epitaxial graphene with thinner top gate dielectrics~\cite{MJHollander:NL11}. 

In this paper, we extend the theory of Adam and co-workers to the more complicated geometry of top-gated SLG, and apply it to study the effect of the top gate on impurity scattering, emphasizing the dependence of the mobility on the top dielectric thickness. Our principal finding is that a thinner top gate leads to a higher carrier mobility because of the stronger image-charge effect in the gate. This agrees with the scaling trend found by Fallahazad and co-workers~\cite{BFallahazad:APL12} as well as Hollander and co-workers~\cite{MJHollander:NL11}. This mobility improvement due to stronger gate screening was also suggested for conventional metal-oxide-semiconductor systems by Gamiz and Fischetti~\cite{FGamiz:APL03}. A general formula for charged impurity scattering in SLG heterostructures is provided.

The outline of the paper is as follows. We derive the Thomas-Fermi screening and the electron-impurity Coulombic interaction, taking into account the boundary conditions imposed by the substrate and the top gate. The matrix element for electron-impurity interaction is calculated and used to compute the momentum relaxation rate which determines the conductivity and carrier mobility. We then compute the carrier mobility for different top dielectrics, gate thickness, and carrier density.

\begin{figure}
\includegraphics[width=2.7in]{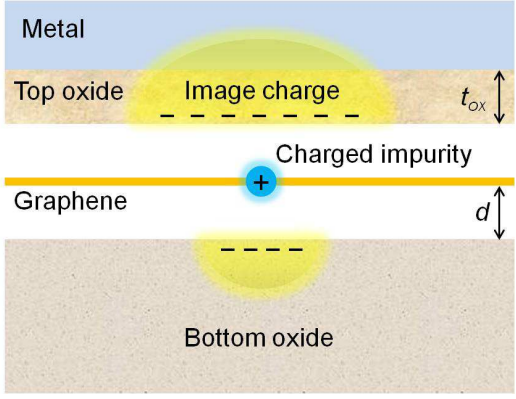}
\caption{Basic model used in our calculation. The SLG is an infinitely thin layer suspended in the middle of the vacuum gap between a semi-infinite substrate and a top oxide layer of thickness $t_{ox}$. The dielectric is capped with metal which we assume to be a perfect conductor. The charged impurity on the SLG has image charges under and above it in the substrate and top gate respectively.}
\label{Fig:GFETStructure}
\end{figure}

{\it Method--}
The model of the double-gated graphene nanostructure consists of a SLG sheet sandwiched between two oxide layers with the interfaces coplanar with the $x$-$y$ plane. The substrate consists of the semi-infinite region $z<0$ while the top oxide spans the region $h\leq z<h+t_{ox}$, where
$h$ is the height of the empty space between the two oxide layers and $t_{ox}$ is the thickness of the top oxide layer. The region $z\geq h+t_{ox}$ is assumed to be composed of a semi-infinite ideal metal. Thus, all field lines terminate at $z=h+t_{ox}$. For simplicity, we assume that the SLG floats at a height of $z=d=h/2$ halfway between the oxide layers. The structure is shown in Fig.~\ref{Fig:GFETStructure}. The dielectric constants of the top and bottom oxides are $\epsilon_{tox}$ and $\epsilon_{box}$ respectively. 

To describe the static dielectric screening of a charged impurity in SLG, we start from the Poisson equation for the \emph{screened} scalar potential $\Phi_{scr}$: 
\begin{equation}
-\nabla^{2} \Phi_{scr}(\mathbf{R},z) 
= 
\frac{1}{\epsilon_{0}} \left[ \rho_{imp}(\mathbf{R},z) + \rho_{scr}(\mathbf{R},z) \right] \ ,
\label{Eq:EffectivePoisson}
\end{equation}
where $\rho_{imp}$ is the impurity charge and $\rho_{scr}$ represents the screening (polarization) charge. Here $\mathbf{R}$ represents the coordinate on the $x$-$y$ plane while $z$ is the coordinate along the perpendicular direction. The integral form of Eq.~(\ref{Eq:EffectivePoisson}) is: 
\begin{equation}
\Phi_{scr}(\mathbf{R},z)
=
\Phi(\mathbf{R},z) + \int d\mathbf{R}' dz' G(\mathbf{R}z,\mathbf{R}'z') \rho_{scr}(\mathbf{R}'z') \ ,
\label{Eq:EffectivePoissonIntegral}
\end{equation}
where $G(\mathbf{R}z,\mathbf{R}'z')$ is the Green function that satisfies the equation $-\nabla^{2}\left[\epsilon(\mathbf{R},z)G(\mathbf{R}z,\mathbf{R}'z')\right]=\delta(\mathbf{R}-\mathbf{R}',z-z')$ with the proper boundary conditions. The bare potential $\Phi(\mathbf{R},z)$ is defined as $\Phi(\mathbf{R},z)=\int d\mathbf{R}'dz'G(\mathbf{R}z,\mathbf{R}'z')\rho_{imp}(\mathbf{R}'z')$. The second term on the right-hand side of Eq.~(\ref{Eq:EffectivePoissonIntegral}) represents the screening charge distribution. The bare and screened potentials can be written in terms of their Fourier components: 
$\Phi(\mathbf{R},z)=\sum_{\mathbf{Q}}\phi_{Q}(z)e^{-i\mathbf{Q}\cdot\mathbf{R}}$ and $\Phi_{scr}(\mathbf{R},z)=\sum_{\mathbf{Q}}\phi_{Q}^{scr}(z)e^{-i\mathbf{Q}\cdot\mathbf{R}}$. The two-dimensional Fourier transform of Eq.~(\ref{Eq:EffectivePoissonIntegral}) yields the following expression for the $z$-dependent part of the Fourier-transformed screened potential: 
\begin{equation}
\phi_{Q}^{scr}(z) 
= 
\phi_{Q}(z)+\int dz'G_{Q}(z,z')\rho_{Q}^{scr}(z') \ ,
\label{Eq:EffectivePoissonIntegralFourier}
\end{equation}
where $\mathbf{Q}$ is the two-dimensional in-plane wave vector. Equation~(\ref{Eq:EffectivePoissonIntegralFourier}) can be solved if we express the polarization charge $\rho_{Q}^{scr}$ as a function of the screened scalar potential. Thus, we express the screening charge as: 
\begin{equation}
\rho_{Q}^{scr}(z) 
= 
e^{2}\Pi(Q)f(z)\phi_{Q}^{scr}(z) \ ,
\label{Eq:ScreeningCharge}
\end{equation}
where $\Pi(Q)$ is the in-plane static polarizability and $f(z)$ governs the polarization charge distribution in the perpendicular direction. We require $f(z)$ to be normalizable to unity, {\em i.e.}, $\int dz\, f(z)=1$. For convenience, we choose $f(z)=\delta(z-d)$. Physically, this implies
that the SLG is idealized as an infinitely thin sheet of polarized charge. Combining Eqs.~(\ref{Eq:EffectivePoissonIntegralFourier}) and (\ref{Eq:ScreeningCharge}), we obtain the expression: 
\begin{equation}
\phi_{Q}^{scr}(z) 
= 
\phi_{Q}(z)+e^{2}\int dz'G_{Q}(z,z')\Pi(Q)f(z')\phi_{Q}^{scr}(z')\ .\label{Eq:EffectivePoissonIntegralFourierFinal}
\end{equation}
After some algebra, Eq.~(\ref{Eq:EffectivePoissonIntegralFourierFinal}) yields:
\[
\phi_{Q}^{scr}(z)=\phi_{Q}(z)+\frac{e^{2}G_{Q}(z,d)\Pi(Q)}{1-e^{2}G_{Q}(d,d)\Pi(Q)}\phi_{Q}(d)\ .
\]
Since we are only concerned about the in-plane scattering potential $\phi_{Q}^{scr}(d)$, we have:
\begin{equation}
\phi_{Q}^{scr}(d) 
= 
\frac{\phi_{Q}(d)}{1-e^{2}G_{Q}(d,d)\Pi(Q)}\ .
\label{Eq:ScreenedScalar}
\end{equation}
Therefore, the generalized expression for the static two-dimensional screening function is $\epsilon(Q)=1-e^{2}G_{Q}(d,d)\Pi(Q)$. In the simplest case of suspended SLG, we have $G_{Q}(d,d)=\frac{1}{2\epsilon_{0}Q}$ and $\Pi(Q)=-\frac{2E_{F}}{\pi\hbar^{2}v_{F}^{2}}$~\cite{SAdam:SSC09}, giving us $\epsilon(Q)=1+\frac{Q_{s}}{Q}$, where $Q_{s}=\frac{e^{2}E_{F}}{\epsilon_{0}\pi\hbar^{2}v_{F}^{2}}$,
$E_{F}$, and $v_{F}$ are the inverse screening length, the Fermi level and the Fermi velocity, respectively. Note that the screening function is determined by its electrostatic environment through $G_{Q}(d,d)$, the electrostatic Green function.

In order to solve Eq.~(\ref{Eq:ScreenedScalar}) and find the screening function $\epsilon(Q)$, we need an explicit expression for the electrostatic Green function $G_{Q}(z,z')$. This Green function
incorporates the details of the electrostatic environment around the SLG. For a source in the empty space between the top and bottom oxide layers, {\em i.e.}, for $0<z'\leq h$,
$G_{Q}(z,z')$ satisfies the equation: 
\[
-\epsilon\left(\frac{\partial^{2}}{\partial z^{2}}-Q^{2}\right)G_{Q}(z,z')=\delta(z-z')\ .
\]
This implies that we can always express $G_{Q}(z,z')$ as a linear combination of $e^{Qz}$ and $e^{-Qz}$, or $\cosh(Qz)$ and $\sinh(Qz)$. The Green function has to satisfy the following continuity conditions at the interfaces $z=0$, $z=h$ and $z=h+t_{ox}$:$G_{Q}(h+t_{ox},z')=\lim_{z\rightarrow-\infty}G_{Q}(z,z')=0$,
$G_{Q}(h^{+},z')=G_{Q}(h^{-},z')$, $G_{Q}(0^{+},z')=G_{Q}(0^{-},z')$, $\epsilon_{tox}\frac{\partial G_{Q}(z=h^{+},z')}{\partial z}=\epsilon_{0}\frac{\partial G_{Q}(z=h^{-},z')}{\partial z}$ and $\epsilon_{0}\frac{\partial G_{Q}(z=0^{+},z')}{\partial z}=\epsilon_{box}\frac{\partial G_{Q}(z=0^{-},z')}{\partial z}$.
In addition, we have the following conditions at $z=z'$:  $G_{Q}(z=z'^{-},z')=G_{Q}(z=z'^{+},z')$
and $\epsilon_{0}\frac{\partial}{\partial z'}G_{Q}(z=z'^{-},z')-\epsilon_{0}\frac{\partial}{\partial z'}G_{Q}(z=z'^{+},z')=1$. 
After some straightforward but lengthly algebra we obtain for $G_{Q}(0<z\leq h,0<z'\leq h)$ the following expression:
\begin{widetext}
\begin{equation}
G_{Q}(z,z') 
=  
\frac{1}{2\epsilon_{0}Q}\Bigg\{ e^{-Q|z-z'|} + \frac{\lambda_{t}\lambda_{b}e^{-2Qh}}{1-\lambda_{t}\lambda_{b}e^{-2Qh}}
\left[ e^{-Q(z-z')} + e^{Q(z-z')} \right] - \frac{\lambda_{t}e^{Q(z+z'-2h)}+\lambda_{b}e^{-Q(z+z')}}{1-\lambda_{t}\lambda_{b}e^{-2Qh}} \Bigg\}
\end{equation}
\end{widetext}
where $\lambda_{t}=\frac{\epsilon_{tox}^{\infty}\coth Qt_{ox}-\epsilon_{0}}{\epsilon_{tox}^{\infty}\coth Qt_{ox}+\epsilon_{0}}$ and $\lambda_{b}=\frac{\epsilon_{box}^{\infty}-\epsilon_{0}}{\epsilon_{box}^{\infty}+\epsilon_{0}}$.
In the region $h\leq z<h+t_{ox}$, making use of the continuity of the Green function and the fact that it has to terminate at $z=h+t_{ox}$, we can easily write the Green function as $G_{Q}(z>h,z')=G(h,z')\frac{\sinh Q(h+t_{ox}-z)}{\sinh Qt_{ox}}$. Similarly, for $z<0$ we have $G_{Q}(z<0,z')=G(0,z')e^{+Qz}$. 

The scattering potential due to a single \emph{bare} point charge on the SLG at the origin satisfies the equation
$-\epsilon_{0}\nabla^{2}\Phi(\mathbf{R},d) = e^{2}\delta(\mathbf{R},d)$.
The corresponding two-dimensional Fourier transform of $\Phi(\mathbf{R},d)$ is $\phi_{Q}(d)=e^{2}G_{Q}(d,d)\ .$ Upon including the screening function from Eq.~(\ref{Eq:ScreenedScalar}), we obtain the \emph{generalized expression} for the screened Coulomb scatterer: 
\begin{equation}
\phi_{Q}^{scr}(d)
=
\frac{e^{2}G_{Q}(d,d)}{1-e^{2}G_{Q}(d,d)\Pi(Q)}\ .
\label{Eq:ScreenedCoulombScatterer}
\end{equation}
The matrix element of the scattering potential of randomly distributed screened impurity charge-centers in SLG is $|\langle V_{s\mathbf{K},s\mathbf{K'}}\rangle|^{2} = |\phi_{|\mathbf{K-\mathbf{K'}|}}^{scr}(d)|^{2}(1+\cos\theta_{\mathbf{K}\mathbf{K'}})/2$, where $\theta_{\mathbf{K}\mathbf{K'}}$ is the angle between the wave vectors $\mathbf{K}$ and $\mathbf{K'}$, and $s$ = +1 (-1) for the conduction (valence) band. The transport scattering rate is given by $1/\tau(E_{s\mathbf{K}})=(n_{imp}/h)\int d\mathbf{K'}|\langle V_{s\mathbf{K},s\mathbf{K'}}\rangle|^{2}[1-\cos\theta_{\mathbf{K}\mathbf{K'}}]\delta(E_{s\mathbf{K}}-E_{s\mathbf{K'}})$, where $n_{imp}$ is the impurity density,
and the conductivity $\sigma$ is approximated by $\sigma=e^{2}v_{F}^{2}D(E_{F})\tau(E_{F})/2$,
where $D(E_{F})$ is the density of states at the Fermi level. The impurity-limited mobility is taken to be~\cite{SAdam:SSC09,SAdam:PNAS07}:
\begin{equation}
\mu=\frac{\sigma}{en}
=
\frac{e^{2}v_{F}^{2}}{2en}D(E_{F})\tau(E_{F}) \ , 
\label{Eq:ImpurityLimitedMobility}
\end{equation}
where $n=E_{F}^{2}/(\pi\hbar^{2}v_{F}^{2})$ is the zero-Kelvin approximation to the carrier density. 

{\it Results and discussion--}
Using Eq.~(\ref{Eq:ImpurityLimitedMobility}), we estimate the impurity-limited mobility of top-gated SLG for different carrier densities, top gate dielectrics, and dielectric thickness. In particular, the dependence of the mobility on $t_{ox}$ is studied in detail. For simplicity, we assume that the doping level is $\geq 10^{12}\mathrm{\ cm^{-2}}$ because charge inhomogeneity (`puddles') is a phenomenon at/near the Dirac point and is not captured in our theory. 

\begin{figure}
\includegraphics[width=2.8in]{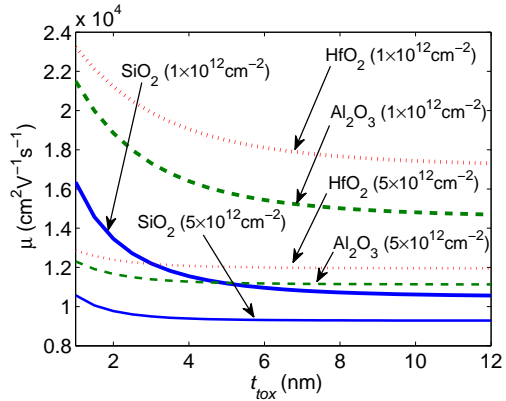}
\caption{Dependence of the carrier mobility ($\mu$) on top gate dielectric thickness ($t_{ox}$) for $\mathrm{SiO_{2}}$, $\mathrm{Al_{2}O_{3}}$, and $\mathrm{HfO_{2}}$. The mobility decreases with increasing $t_{ox}$. This decrease is greater at $n = 10^{12}\ \mathrm{cm^{-2}}$ than at $n = 5\times 10^{12}\ \mathrm{cm^{-2}}$.}
\label{Fig:ThicknessDependence}
\end{figure}

We consider three top gate dielectrics, $\mathrm{SiO_{2}}$ ($\kappa$=3.9), $\mathrm{Al_{2}O_{3}}$ ($\kappa$=12.5)~\cite{BFallahazad:APL12} and  $\mathrm{HfO_{2}}$($\kappa$=22.0)~\cite{BFallahazad:APL10}. The bottom dielectric is assumed to be $\mathrm{SiO_{2}}$. $t_{ox}$ is varied from 1 to 12 nm. The electron mobility is calculated for two values of the carrier density, $\mathrm{1\times10^{12}\ cm^{-2}}$
and $\mathrm{5\times10^{12}\ cm^{-2}}$. The results are shown in Fig.~\ref{Fig:ThicknessDependence}.
We set the impurity density to be $n_{imp}=0.5\times10^{12}\ \mathrm{cm^{-2}}$.
In Fig.~\ref{Fig:ThicknessDependence}, we notice that the dependence of the mobility on $t_{ox}$ is stronger for $n=\mathrm{1\times10^{12}\ cm^{-2}}$ than for $n=\mathrm{5\times 10^{12}\ \text{cm}^{-2}}$. As expected, the higher-$\kappa$ oxides yield a larger mobility because the higher effective permittivity weakens Coulombic interactions (electron-electron and electron-impurity) in graphene.
Although the weaker electron-electron interaction results in less screening charge around the impurity, the bare charge on the impurity is reduced to a greater extent, resulting in a overall diminished screened impurity charge.

\begin{figure}
\includegraphics[width=2.8in]{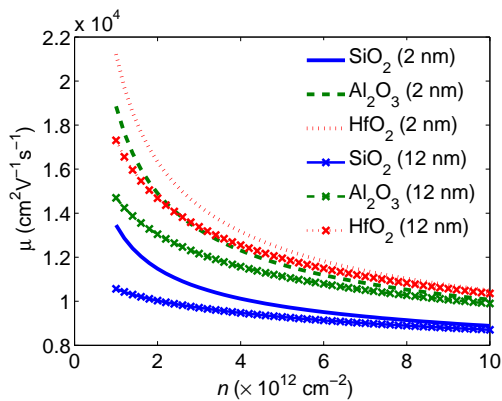}

\caption{Dependence of the mobility on carrier density. The mobility decreases
with carrier density, and the extent of the decrease is greater with
smaller top gate dielectric thickness.}

\label{Fig:CarrierDensDependence}
\end{figure}

We also find that the mobility increases with decreasing $t_{ox}$, in qualitative agreement with the experimental results in Ref.~\cite{BFallahazad:APL12}.  
In our simulations, this increase in mobility can be substantial. For example, in Fig.~\ref{Fig:ThicknessDependence} at $n = 10^{12}\ \mathrm{cm^{-2}}$, the mobility for a 1.0 nm $\mathrm{SiO_{2}}$ top gate is almost 60 percent higher than that for a 12.0 nm top gate. 
This scaling trend has been attributed to the increase in defect concentration with greater $t_{ox}$~\cite{BFallahazad:APL12}. However, our model also reproduces this trend because with a decreasing $t_{ox}$ the image `anti-charge' at the metal-dielectric interface gets closer to the impurity charge and weakens its scattering potential. This implies that ultra-thin top gates can be used to compensate for large charged impurity densities. 
The carrier-density dependence of the mobility is calculated for different top gate dielectrics at $t_{ox}$= 2 nm and 12 nm. The carrier density is varied from $10^{12}$ to $10^{13}\ \mathrm{cm^{-2}}$. The results are shown in Fig.~\ref{Fig:CarrierDensDependence}.
In general, the mobility decreases with increasing carrier density, in contrast to the results in Refs.~\cite{SAdam:SSC09,SAdam:PNAS07} which suggest that mobility should be independent of carrier density. This decrease is more acute in the 2 nm case than in the 12 nm case and for higher-$\kappa$ oxides. 
To understand why, we look at Eq.~(\ref{Eq:ImpurityLimitedMobility}). In suspended SLG, the Fermi level $E_F$ is proportional to $n^{1/2}$. The scattering time $\tau(E_F)$ is also proportional to $n^{1/2}$ because $G_{Q}(d,d)$ in Eq.~(\ref{Eq:ScreenedCoulombScatterer}) scales as $1/Q$. Thus, $\mu$ is independent of $n$. On the other hand, in top-gated SLG, $G_{Q}(d,d)$ is almost constant over a large range of $Q$ values, and hence, $\tau(E_F)$ is approximately independent of $n$ at low carrier densities. Thus, $\mu$ scales roughly as $n^{-1/2}$. 
Thus, the trend shown in Fig.~\ref{Fig:CarrierDensDependence} should be viewed as low-density/thin-oxide enhancement of the mobility above its gate-unscreened value thanks to the screening effect of the image charges induced in the gate by the impurities.

The improvement of mobility with a thinner top oxide layer highlights the compensating effect of having a metal layer in close proximity to SLG. It has been suggested that voltage fluctuations induced by charged impurities in the substrate lead to local electrostatic
doping and are responsible for the formation of `puddles' in graphene near the charge neutral point~\cite{JMartin:NaturePhysics07,VMGalitski:PRB07}. We propose that doping inhomogeneities can be `smoothened out' through the use of an ultrathin high-$\kappa$ top-gate structure since the image
charges can partially neutralize the effects of the charged impurities. Unintentional doping from the substrate can also be reduced, leading to an improvement of carrier transport in top-gated SLG.

{\it Conclusions--}
We have studied charged impurity scattering and static screening in top-gated SLG nanostructures. The image charge effect dampens Coulombic interaction, resulting in weaker unscreened charged impurity scattering and static screening. 
The carrier mobility is found to increase with decreasing top gate dielectric thickness and decreasing carrier density. Our theory suggests that if charged impurity scattering is a significant factor in limiting electronic transport in SLG, then having a top metal gate with an ultrathin dielectric can compensate for the effect of the charged impurities. This offers a simple and practical design strategy to improve the impurity-limited mobility in SLG heterostructures.

{\it Acknowledgement--}
We gratefully acknowledge the support provided by Texas Instrument, the Semiconductor Research Corporation (SRC), the Microelectronics Advanced Research Corporation (MARCO), the Focus Center Research Project (FCRP) for Materials, Structures and Devices (MSD), and Samsung Electronics Ltd.

\bibliographystyle{apsrev4-1}
\bibliography{TopGateImpurityScattering_v25June2012}

\end{document}